\documentclass[superscriptaddress,twocolumn,prl]{revtex4-1}

\usepackage{amssymb,amsmath,graphicx}
\usepackage{color}
\draft 

\begin{document}

\title{Dielectric geometric phase optical elements from femtosecond direct laser writing} 

\author{Xuewen Wang}
\email[]{Email: xuewenwang@swin.edu.au}
\affiliation{Centre of Micro-Photonics, Swinburne University of
Technology, Hawthorn, VIC 3122, Australia}

\author{Aleksandr Kuchmizhak}
\affiliation{Centre of Micro-Photonics, Swinburne University of
Technology, Hawthorn, VIC 3122, Australia} \affiliation{School of
Natural Sciences, Far Eastern Federal University (FEFU), 8
Sukhanova str., Vladivostok 690041, Russia} \affiliation{Institute
of Automation and Control Processes of FEB RAS, 5 Radio Str.,
690041 Vladivostok, Russia}

\author{Etienne Brasselet}
\affiliation{Universit\'{e} de Bordeaux, LOMA, UMR 5798, F-33400
Talence, France}
\affiliation{CNRS, LOMA, UMR 5798, F-33400
Talence, France}

\author{Saulius Juodkazis}
\email[]{Email: sjuodkazis@swin.edu.au}
\affiliation{Centre of Micro-Photonics, Swinburne University of
Technology, Hawthorn, VIC 3122, Australia} \affiliation{Melbourne
Centre for Nanofabrication, ANFF, 151 Wellington Road, Clayton,
VIC 3168, Australia}

\date{\today}

\begin{abstract}
We propose to use femtosecond direct laser writing technique to
realize dielectric optical elements from photo-resist materials
for the generation of structured light from purely geometrical
phase transformations. This is illustrated by the fabrication and
characterization of spin-to-orbital optical angular momentum
couplers generating optical vortices of topological charge from 1
to 20. In addition, the technique is scalable and allows obtaining
microscopic to  macroscopic flat optics. These results thus
demonstrate that direct 3D photopolymerization technology
qualifies for the realization of spin-controlled geometric phase
optical elements.
\end{abstract}


\maketitle

During the last two decades, the concept of geometric phase
optical elements \cite{bhandari_physrep_1997} established a new
standard in the realization of smart flat optics. The
characteristic of such optical elements is the capability to
impart an arbitrary phase profile to an incident light field by
purely geometrical means. This is made possible by preparing
space-variant optically anisotropic materials. In practice, this
is achieved by preparing a slab with in-plane effective optical
axis whose orientation angle is spatially modulated, say
$\psi(x,y)$. An essential feature is the fact that the optical
functionality encoded in the spatial distribution of the optical
axis is controlled by the polarization state of the light. Indeed,
considering the simplest situation of a transparent slab having a
birefringent phase retardation of $\pi$, an incident circularly
polarized light field impinging at normal incidence (hence along
the $z$ axis) emerges as a contra-circularly polarized field
endowed with a space-variant Pancharatnam-Berry phase
\cite{pancharatnam_pia_56, berry_jmo_1987} $\Phi(x,y) =
2\sigma\psi(x,y)$ where $\sigma \pm 1$ refers to the helicity of
the incident light.

Experimentally, the realization of geometric phase optical
elements has started 15 years ago by designing space-variant
subwavelength gratings made from metal~\cite{biener_ol_2001} and
semiconductor~\cite{biener_ol_2002}, though initially restricted
to mid-infrared domain. The use of dielectric materials  emerged a
few years after by implementing liquid crystals with inhomogeneous
in-plane molecular orientation~\cite{marrucci_apl_06}, thus
providing optical elements operating in the visible domain.
Nowadays, photo-alignment techniques allows obtaining arbitrary
phase profiles from patterned liquid crystal
slabs~\cite{kim_optica_2015}. Still, several other techniques have
been explored in the recent years towards the realization of
dielectric geometric phase optical elements from structured
solid-state materials with great application potential owing to
enhanced lifetime and damage threshold. One can mention
femtosecond direct laser writing (DLW) in
glasses~\cite{beresna_apl_11}, which however suffers from large
scattering losses at visible frequencies, and electron beam
lithography of silicon~\cite{lin_science_14, desiatov_oe_15, Kruk}
and titanium oxide~\cite{devlin_pnas_2016}.

\begin{figure}[b!]
\centering{\includegraphics[width=0.87\linewidth]{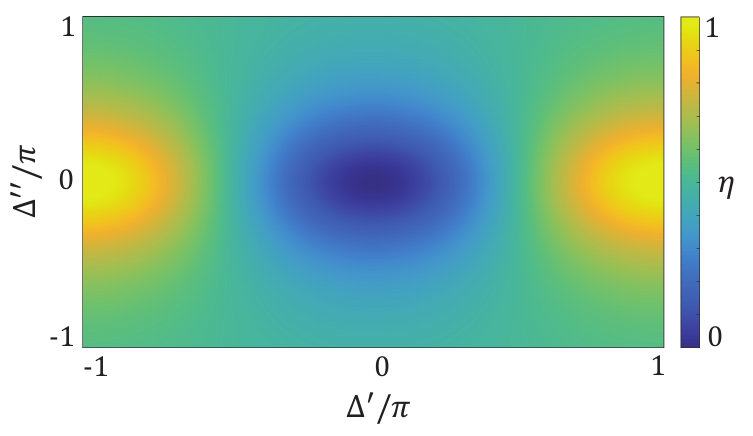}}
\caption{Purity $\eta$ as a function of the real ($\Delta'$) and
imaginary ($\Delta''$) parts of the complex birefringent phase
retardation. Positive and negative values for $(\Delta',\Delta'')$
refer to positive and negative uniaxial behavior, respectively.}
\label{fig_eta}
\end{figure}

In practice, dielectrics offer the advantage of possible
transparency over very a large spectral range, which favors the
elaboration of high-transmission devices. On the other hand,
high-refractive indices enable optimal cross-polarization
conversion with thin layers with respect to wavelength. Besides
the average refractive index mismatch between the external and
structured media and the average attenuation that both affect the
overall transmission, the dichroism (i.e., the anisotropy of the
imaginary part of the complex dielectric permittivity tensor) has
a direct influence on the helicity transformation $\sigma \to
-\sigma$. More precisely, the purity parameter $\eta$ defined as
the fraction of the output power that corresponds to
helicity-flipped field experiencing the Pancharatnam-Berry phase
can be expressed as \cite{DavitPhD} $\eta = \left(1 -
\cos\Delta'/\cosh\Delta''\right)/2$ where $\Delta'=kh(n_\parallel'
- n_\perp')$ and $\Delta''=kh(n_\parallel'' - n_\perp'')$, with
$k$ the wavevector in vacuum, $h$ the thickness of the element,
$n' + in''$ its complex refractive index and $(\parallel,\perp)$
referring to direction parallel and perpendicular to the local
effective optical axis. Interestingly, the dichroism may enhance
or reduce the purity depending on the real birefringent phase
retardation, as illustrated in Fig.~\ref{fig_eta}.

Here we propose an unexplored yet approach for the fabrication of
dielectric geometric phase optical elements based on femtosecond
DLW of photo-resists, although the technology is now mature
\cite{13pr1}. An asset of this approach is that it is
easy-to-implement while the realization of macroscopic dimensions
is possible. This contrasts to currently employed nanofabrication
techniques based on electron beam lithography, focused ion
milling, and atomic layer deposition of dielectric layers that
remain the privilege of cleanroom facilities and require
high-level technical support. Moreover, the inherent
three-dimensional structuring capabilities of proposed approach
allows considering the fabrication of dielectric devices on curved
and flexible substrates~\cite{16le16133} while the structured
material itself can be reconfigurable under external forcing for
instance by using elastomers~\cite{hbsun}. In present case, we
choose the hybrid (20\% inorganic, 80\% organic) photo-resist
SZ2080 whose refractive index is $n^{'} + in^{"} = 1.474 + i0.08$
over the $500-800$~nm wavelength range~\cite{2015Baldacchini}.
Such a material has a low shrinkage, high optical transmissivity,
and is widely used for micro-optical
elements~\cite{Farsari,10apl211108,16oe16988,16aom}. In principle,
highly pure geometric phase optical elements can thus be formally
obtained under appropriate optimization of the designed structure
following above discussion on the parameter $\eta$.

\begin{figure}[b!]
\centering{\includegraphics[width=\linewidth]{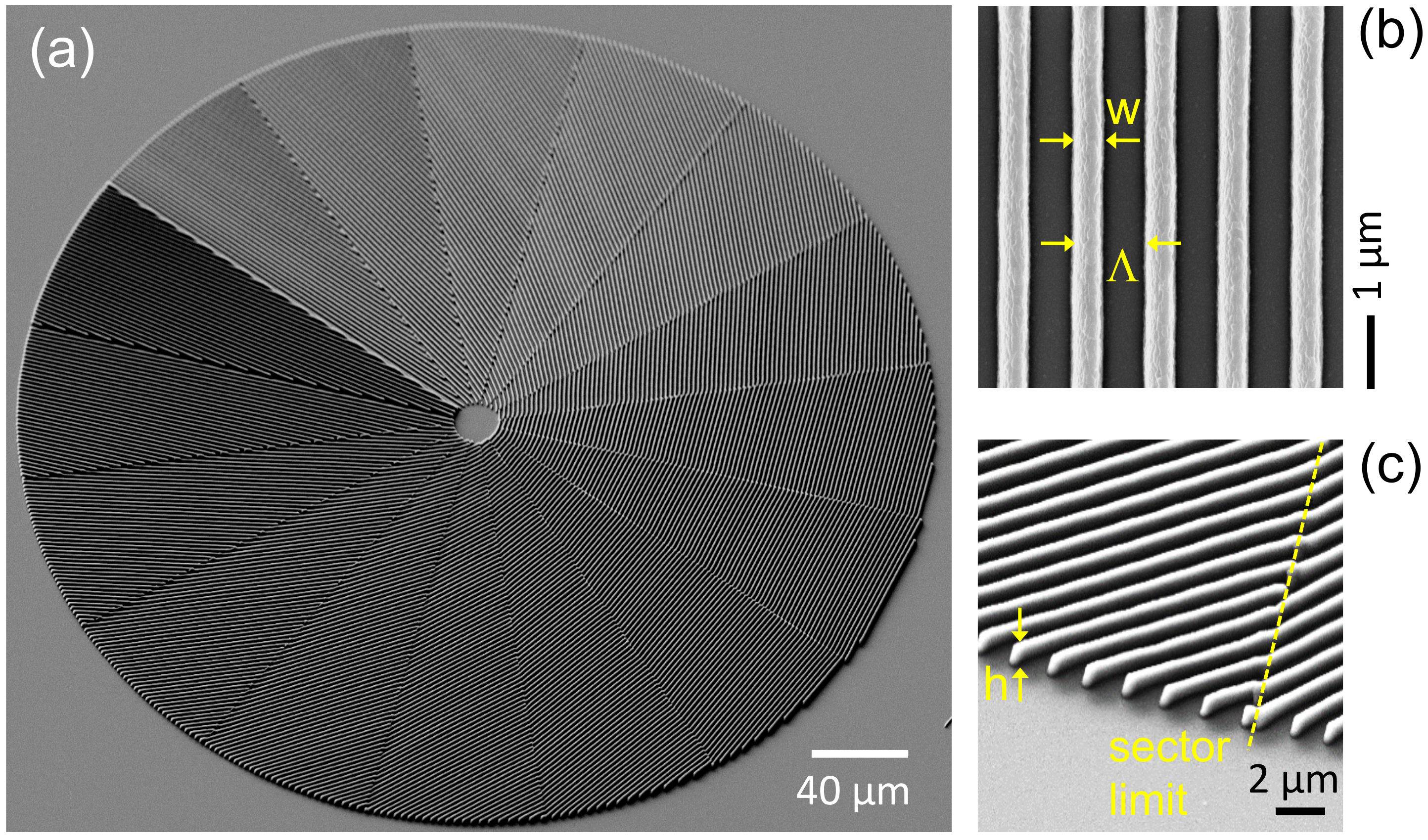}} \caption{(a)
45$^\circ$-slanted SEM image of a 16-step $\frac{1}{2}$-plate of
diameter 200~$\mu$m. Note that surface charging is altering the
imaging contrast. (b) Top-view SEM image of the local
photo-polymerized grating characterized by the filling factor
$w/\Lambda \simeq 0.3$. (c) SEM image at the rim of the element,
where $h=1~\mu$m is the height of the structure.} \label{fig:3d}
\end{figure}

Without lack of generality, we restrict our demonstration to the
realization and characterization of spin-to-orbital optical
angular momentum converters. Such elements correspond to
azimuthally varying optical axis orientation of the form $\psi =
q\phi$ ($q$ half-integer and $\phi$ the polar angle in cylindrical
coordinates) with, ideally, uniform real birefringent phase
retardation of $\pi$ \cite{Marrucci2006}. In turn, a spin-orbit
coupler transforms an incident field with helicity $\sigma$ (that
is associated with spin angular momentum $\sigma\hbar$ per photon)
into an helicity flipped field endowed with a spatial distribution
of the phase of the form $\Phi(\phi) = 2\sigma q \phi$ (that is
associated with orbital angular momentum $2\sigma q\hbar$ per
photon). The choice of such a design to test our approach is
motivated by the wide range of applications in classical and
quantum optics of these so-called $q$-plates \cite{Marrucci2011}
that have become a prototypical benchmark for geometric phase
optical elements.

The DLW experimental platform basically consists of a regenerative
amplified Yb:KGW based femtosecond fs-laser system (Pharos, Light
Conversion Ltd.) operating at the second harmonic wavelength of
515~nm with pulse duration of 230~fs and 200~kHz repetition rate.
The laser beam is focused with an oil-immersion objective lens of
numerical aperture $NA = 1.42$ (Olympus) onto the interface of a
cover glass on which the dielectric photo-resist doped with
1\%wt. 4,4'-bis-diethylaminobenzophenone as a photoinitiator is
drop-casted and dried under room temperature for 12~h before laser
writing. The pulse energy after the objective is set to 0.12~nJ at
the scanning speed of 0.1~mm/s. After fabrication, the samples
were developed in a methyl-isobutyl-ketone and isopropanol 1:2
solution for 10~min followed by a 1~min acetone bath and 10~min
rinse in isopropanol. Then, the structures were dried on a
hotplate at 50$^\circ$C for 10~min. Finally, a 5-nm-thick film of
titanium was sputtered for structural characterization by scanning
electron microscopy (SEM).

\begin{figure}[t!]
\centering{\includegraphics[width=\linewidth]{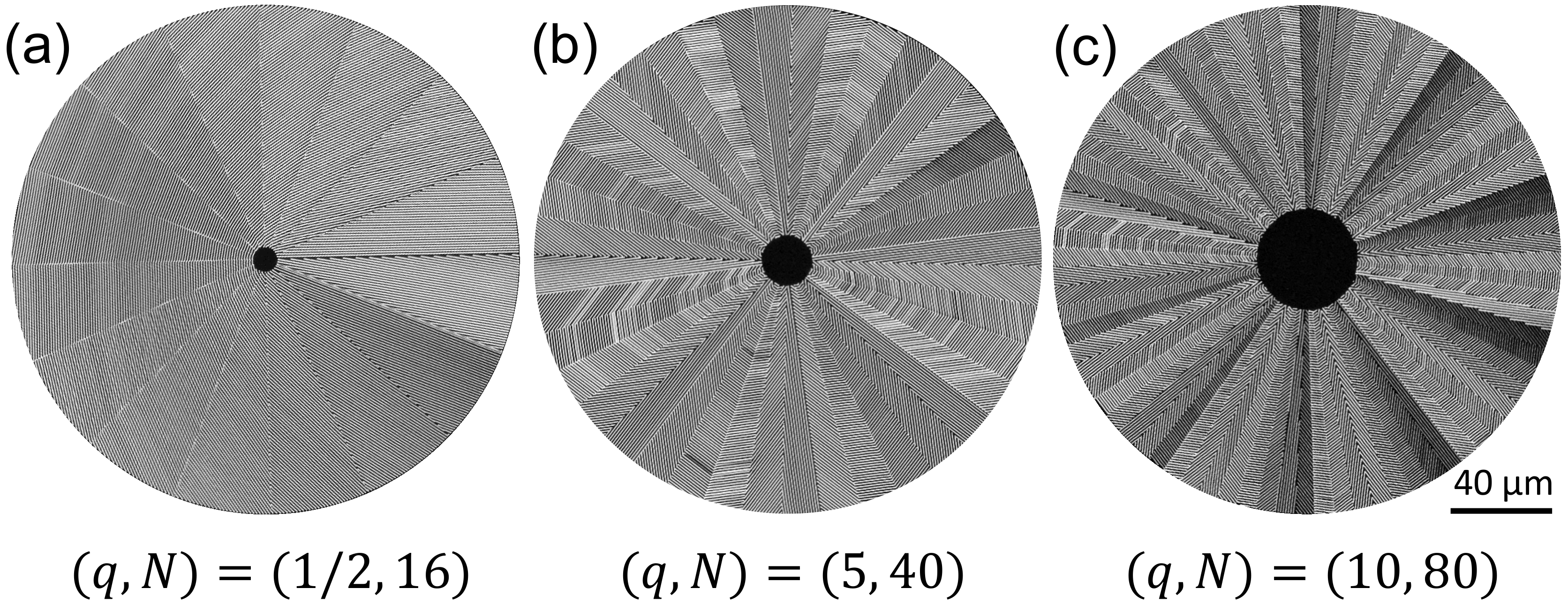}}
\caption{Top-view SEM images of various $N$-step $q$-plates
enabling spin-orbit optical vortex generation of topological
charge $\ell=\pm1$ (a), $\ell=\pm10$ (b), and $\ell=\pm20$ (c).
Each element has diameter of 200~$\mu$m and a purposely inner
unstructured disk of diameter $d=10~\mu$m (a), 20~$\mu$m (b), and
40~$\mu$m (c). Indeed, structuring the central part is eventually
not useful atnwhen the grating pitch $\Lambda$ is of the order of
$d\pi/N$, which gives $d \simeq N\Lambda/\pi=5$, 13 and 25 $\mu$m,
respectively.} \label{fig:structures}
\end{figure}

A $N$-step discrete design is chosen for the space-variant grating
structure associated with a pitch $\Lambda \simeq 1~\mu$m and
filling factor defined by the width-to-period ratio of the
gratings $w/\Lambda \simeq 0.3$, as illustrated in
Fig.~\ref{fig:3d} for $q=1/2$ and $N=16$. High-charge elements
have also been fabricated, as shown on Fig.~\ref{fig:structures}
for $(q,N) = (1/2,16)$, $(5,40)$ and $(10,80)$ that correspond to
fabrication time around 10, 35 and 40~min, respectively, which is
acceptable for industrial DLW. In practice, patterns of lower
complexity and centimeter square area can be made in several hours
using faster scanning and higher laser repetition rate.

\begin{figure}[b!]
\centering{\includegraphics[width=\linewidth]{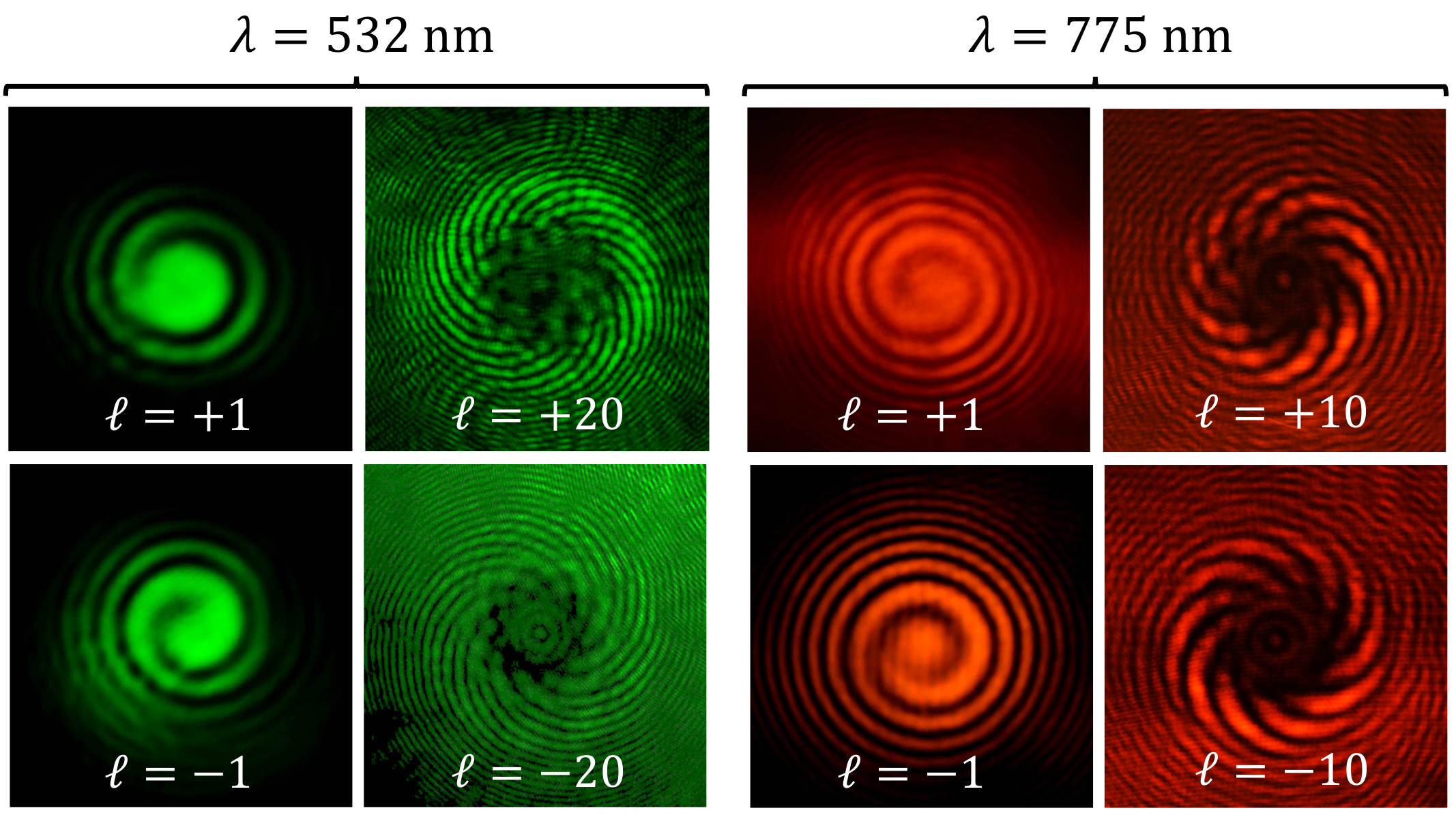}} \caption{
Single-beam interference patterns resulting from spin-orbit
optical vortex generation from the $q$-plates shown in
Fig.~\ref{fig:structures}). The generation of on-axis optical
phase singularity of topological charge $\ell=2\sigma q$ is
identified from the number (given by $|\ell|$) and the handedness
(given by the sign of $\ell$) of the spiraling patterns. }
\label{fig:donuts}
\end{figure}

The optical characterization of the fabricated spin-to-orbital
couplers is made by inspecting the spiraling phase profile
imprinted by the structures to the contra-circularly polarized
output field component. In practice this is made in a
straightforward manner by illuminating the sample by a
$\sigma$-polarized collimated beam of typical diameter 1~mm and
subsequent polarization imaging the intensity distribution of the
field that emerges from the sample. Indeed, the diffraction of
light on the finite size $q$-plate having central unstructured
area (see Fig.~\ref{fig:structures}) leads to ``single-beam
interferometry" by providing coaxial overlap between the two
circularly polarized output field components, hence without need
of an external reference beam. This leads to spiraling fringes
patterns whose contrast is optimized by adjusting the polarization
state on which the total field is projected. This is made by
placing an achromatic quarter-waveplate followed by a polarizer
after the sample and adjusting their relative orientation. The
results are illustrated in Fig.~\ref{fig:donuts} for the
structures shown in Fig.~\ref{fig:structures} at two different
wavelengths (532~nm and 775~nm) and for incident helicity
$\sigma=\pm1$. As expected, $2|\sigma q|$-arm spiraling patterns
with helicity-dependent handedness are observed, which
demonstrates the generation of optical vortex beams associated
with an optical phase singularity of topological charge
$\ell=2\sigma q$.

\begin{figure}[t!]
\centering{\includegraphics[width=\linewidth]{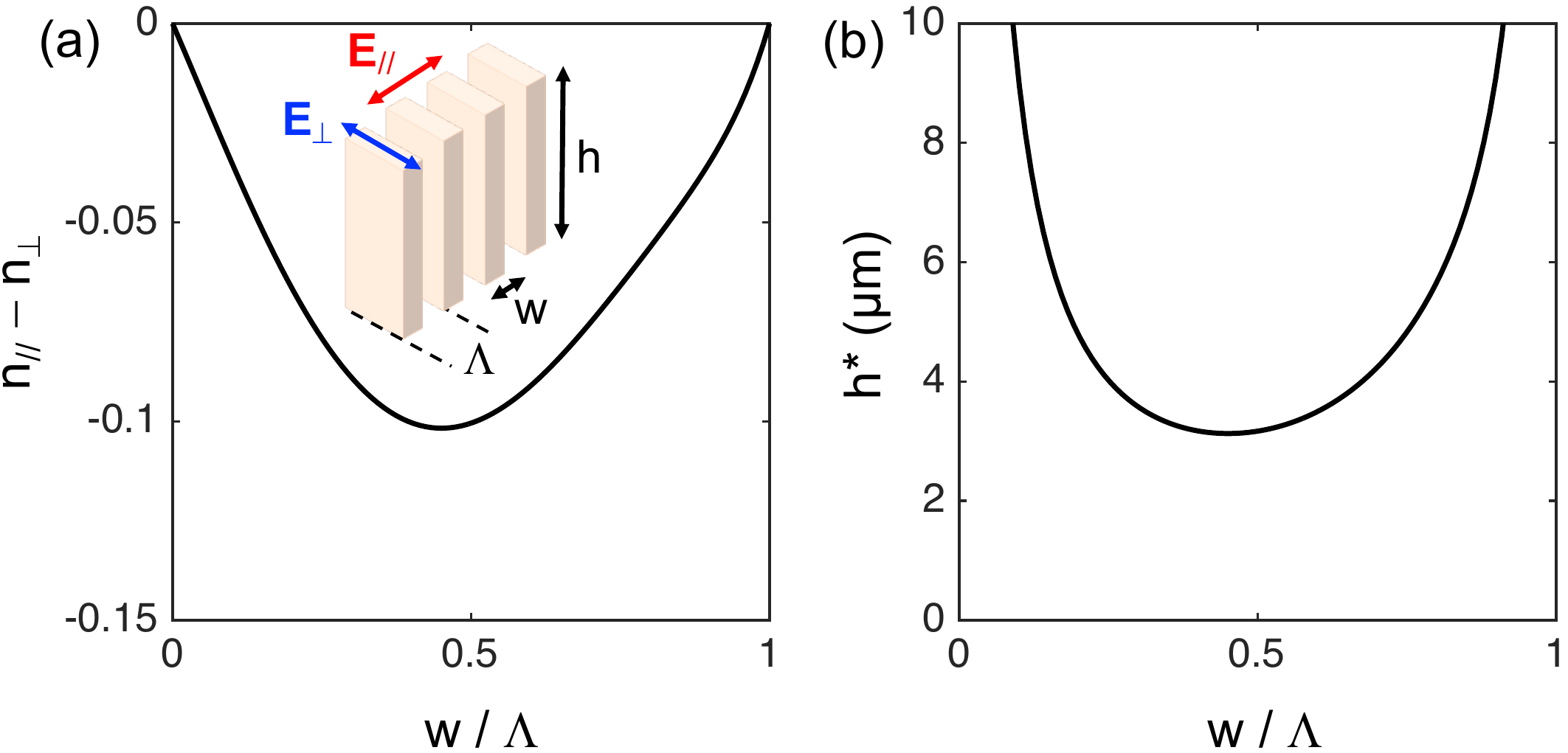}} \caption{(a)
Optical form anisotropy calculated from second order effective
medium theory vs filling factor for $\Lambda=\lambda/2=500~\mu$m
and $n=1.5$, see text for details. Inset: sketch of subwavelength
grating design with corresponding definitions of electric field
components parallel and perpendicular to the grating wavevector.
(b) Height of the structure giving optimal form birefringence as a
function of the filling factor. }\label{fig:emt}
\end{figure}

On the other hand, the performance of the photo-polymerized
geometric phase optical elements is experimentally evaluated to a
few percents. Although such a modest value does not compromise the
proof-of-principle that femtosecond DLW of photo-polymeric
materials is an approach that is worth to explore further, this
invites to consider how to optimize it. For this purpose, an
option is to calculate the complex form birefringence phase
retardation $\Delta = \Delta' + i\Delta''$, which can be done by
using effective medium theories or brute force FDTD simulations.
In practice, inherent resolution of the DLW technique is
restricted to $\Lambda \lesssim \lambda$ in the visible range, see
for instance Ref.~\cite{MM_OL_2014} reporting on grating pitch
$\Lambda=300-400$~nm using standard DLW while twice smaller values
are accessible to super-resolution DLW techniques
\cite{wegener_review_2013}. In turn, second-order effective medium
theory appears as a relevant, yet simple, analytical tool to
design an optimal structure if grating pitch is small enough.
Indeed, the latter approach is typically considered valid up to
$\Lambda \simeq \lambda/2$~\cite{Emoto}. Although chosen
parameters for present experimental demonstration are obviously
not optimal, it is instructive to have a look on expected
parameters enabling optimal performances. This is done by applying
second-order effective medium theory, see Eqs.~(1) and (2) of
Ref.~\citenum{Emoto}, assuming for the sake of illustration
$\Lambda = \lambda/2$ with $\Lambda=500$~nm and $n=1.5$. One gets
an optical anisotropy that depends on the filling factor according
to Fig.~\ref{fig:emt}(a) where $n_\parallel$ and $n_\perp$ are the
effective refractive indices parallel and perpendicular to the
grating wavevector lying in the plane of the structure. Then, the
optimal height $h^*$ satisfying the optimal birefringent phase
retardation $\Delta=\pi$ is evaluated from
$h^*=\lambda/(\pi|dn|)$, see Fig.~\ref{fig:emt}(b). The latter
optimized structure height is in the range $3.1-3.5~\mu$m for
filling factor in the range $0.3-0.6$, which implies design
flexibility. However, the ability to fabricate polymerized line
with aspect ratio $h/w \sim 5$ should not be eluded and certainly
deserves further work to validate robust processing solutions,
since structures with aspect ratio of 18 have been recovered by a
wet bath development of a negative resist~\cite{05apa1583} and
even more delicate structures can be retrieved by avoiding the
capillary forces via a critical point drying
process~\cite{16le16133}. In practice, such conclusions regarding
the optimal purity still applies qualitatively in case of moderate
dichroic losses (say $|\Delta'/\Delta''|<0.1$) as shown in
Fig.~\ref{fig_eta}.

\begin{figure}[t!]
\centering{\includegraphics[width=\linewidth]{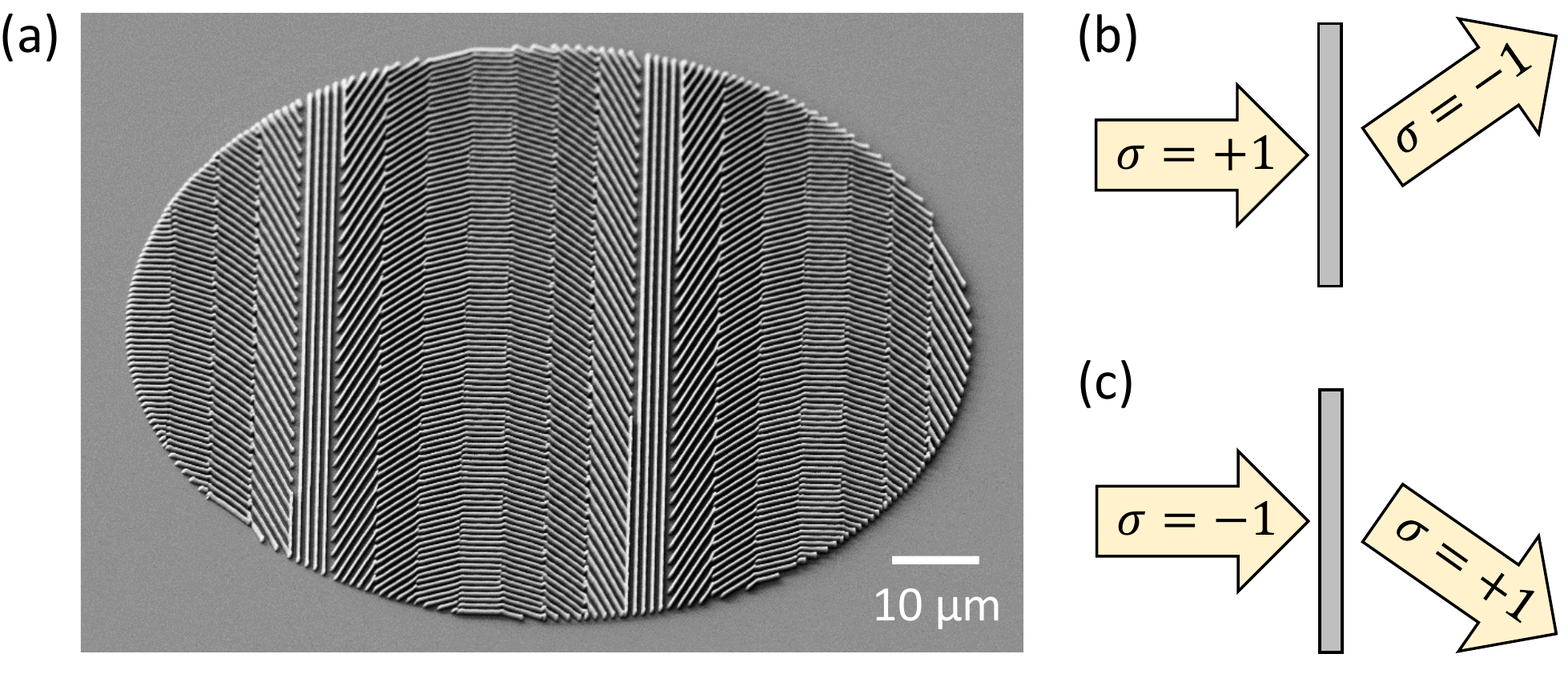}} \caption{(a)
45$^\circ$-slanted SEM image of a discretized optical spin
splitter whose operation principle is illustrated on panels (b)
and (c). }\label{fig:spinsplitter}
\end{figure}

As said above, geometric phase optical elements are not restricted
to spin-to-orbital angular momentum couplers and DLW technology is
versatile. This is illustrated in Fig.~\ref{fig:spinsplitter}(a)
that shows SEM image of a discretized optical spin splitter
enabling helicity dependent redirection of light. Such a device
consists of a one-dimensional grating orientation angle
distribution of the form $\psi = \kappa x$ leading to a tilt of an
incident phase front of the contra-circular output component, as
depicted in Figs.~\ref{fig:spinsplitter}(a) and
\ref{fig:spinsplitter}(c) for $\Delta=\pi$.

In summary, we proposed a novel technique to fabricate geometric
phase optical elements using femtosecond direct laser writing of
photo-resists. The approach is demonstrated by realizing
spin-orbit optical vortex generators of topological charge from 1
to 20 and optical spin splitters.  Such space-variant
form-birefringent structures basically work over a broad spectral
range, though at the expense of overall efficient since the
optimal birefringent phase retardation condition is satisfied only
for well-defined frequencies. In other words, the very same design
principle is applicable for IR and THz spectral ranges, where
application potential is likely for sensing applications. More
generally, by enriching the geometric phase optical elements
toolbox with a nowadays matured technology, our results
contributes to the further developments of spin-orbit photonics.

\vspace{0.25cm}{\small{We acknowledge Workshop of Photonics R\&D.
Ltd. for the laser fabrication setup acquired via a collaborative
grant. We are grateful to Mangirdas Malinauskas for discussions on
laser printing conditions. Partial support by the NATO grant
SPS-985048, the Australian Research Council DP130101205 and
DP170100131 Discovery projects is also acknowledged.}}

%

\end{document}